\documentclass[pra,twocolumn,aps,groupedaddress,showpacs]{revtex4-1}
\usepackage{amsmath}
\usepackage{amssymb}
\usepackage{lineno}
\usepackage{graphicx}
\usepackage[usenames,dvipsnames]{color}
\usepackage{bm}
\usepackage{times}
\usepackage{hyperref}
\usepackage{float}
\usepackage{lipsum}
\usepackage{color}
\usepackage[dvipsnames]{xcolor}
\hypersetup{
  colorlinks=true,
  citecolor=blue,
  linkcolor=blue,
  urlcolor=blue}

\definecolor{orange(colorwheel)}{rgb}{1.0, 0.5, 0.0}
\definecolor{maroon(html/css)}{rgb}{0.5, 0.0, 0.0}
\definecolor{ao(english)}{rgb}{0.0, 0.5, 0.0}
\definecolor{armygreen}{rgb}{0.29, 0.33, 0.13}
\definecolor{britishracinggreen}{rgb}{0.0, 0.26, 0.15}

\begin{document}

\title{Collective modes in multicomponent condensates with anisotropy}
\author{Sukla Pal}
\author{Arko Roy}
\altaffiliation{Present address: 
Max-Planck-Institut f\"{u}r Physik komplexer Systeme,
N\"{o}thnitzer Stra\ss e 38, D-01187 Dresden, Germany.}
\author{D. Angom}
\affiliation{Physical Research Laboratory,
              Ahmedabad-380009, Gujarat,
             India}

\begin{abstract}
We report the effects of anisotropy in the confining potential on two 
component Bose-Einstein condensates (TBECs) through the  properties of 
the low energy quasiparticle excitations. Starting from generalized Gross 
Pitaevskii equation, we obtain the Bogoliubov de-Gennes (BdG) equation for
TBECs using the Hartree-Fock-Bogoliubov (HFB) theory. Based on this theory, we 
present the influence of radial anisotropy on TBECs in the immiscible or the
phase-separated domain. In particular, the TBECs of $^{85}$Rb~-$^{87}$Rb and
$^{133}$Cs~-$^{87}$Rb TBECs are chosen as specific examples of the two possible
interface geometries, shell-structured and side by side, in the immiscible 
domain. We also show that the dispersion relation for the TBEC shell-structured
interface has two branches, and anisotropy modifies the energy scale and
structure of the two branches.

\end{abstract}


\maketitle

\section{Introduction}

The physics of trapped ultra-cold atoms, specifically the Bose Einstein 
condensate (BEC), is replete with novel phenomena emerging from the atom-atom
interactions, trap geometry, quantum and thermal fluctuations, topological
defects, spatial dimensions and so on. In particular, the geometry of the 
confining potential has a strong impact on the density profiles of the single 
component as well as on the multi-component BECs. An example is, in a three 
dimensional (3D) harmonic potential, if the axial frequency ($\omega_z$) is 
much larger than the radial frequencies
($\omega_x = \omega_y=\omega_{\perp}$), the condensate is effectively 2D
with pancake shaped density profile. The symmetric radial frequencies is an 
ideal situation, and in experiments, there are deviations from symmetry due 
to practical limitations of the various components. For example, the 
pioneering experiments on BECs \cite{davis_95, anderson_95} were done in 
anisotropic trap. Thus, it is of practical importance to consider anisotropic 
radial confinement ($\omega_x \ne \omega_y $) and examine the deviations from 
the symmetric case. With this consideration, in the present work, we study the 
effects of radial anisotropy arising from the confining potential at zero 
temperature. An immediate consequence of the anisotropy is the change of 
interface geometry of the multicomponent condensates at phase-separation. 

The inter-atomic interactions play dramatic role in low dimensional,
quasi 1D and 2D, BECs of trapped atomic gases. With quasi low dimensional BECs, there are excitations unique to each which do not have an analogue in
the higher or lower dimensional BECs. In our previous works, we characterized
the low energy excitation modes for quasi 1D 
\cite{roy_15} and quasi 2D \cite{roy_16} BEC. Apart from this, in the 
TBECs the miscible-immiscible phase transition
has strong dimensional dependence. At zero temperature, the condition of
phase separation under Thomas-Fermi (TF) approximation is given by the
inequality $U^2_{12} > U_{11}U_{22}$ \cite{timmermans_98,pu_98,ho_96}. 

In this work, we report the transition from circular to planar
interface with increased  anisotropy \cite{gautam_10a, mertes_07}
in the TBECs
with shell structured density configuration in the immiscible domain. 
We also demonstrate that the transformation in the interface geometry 
leads to a change in the low energy excitation modes. As a representative 
example of the shell structured interface we choose $^{85}$Rb~-$^{87}$Rb TBEC 
which is experimentally well studied \cite{papp_06, papp_08}. The other
density configuration of a TBEC in the immiscible domain is the side by side 
density profiles. For this case as well, we show that the presence of 
anisotropy leads to distinct structures of the density profiles in the
immiscible domain. This is accompanied by the changes in the low energy 
BdG spectrum and structure of the quasiparticle amplitudes. We consider the
$^{133}$Cs~-$^{87}$Rb TBEC \cite{mccarron_11} as a representative example 
for this case.

The dispersion relation is the key to understand how the TBEC responds to
external perturbations. So, to relate theoretical investigations with 
experimental findings, it is essential to determine how the dispersion
relation of TBECs change with anisotropy. In this context analysing the 
observations from the Bragg 
Bogoliubov spectrum which in turn necessitates knowledge of dispersion
relations. Theoretically, dispersion relation for BEC has 
been investigated in the analytic framework \cite{pethick_08}. In addition, 
there are several numerical computations of the dispersion relations in 
finite sized BEC \cite{wilson_10, bisset_13, blakie_2012} in presence of 
roton like spectrum. The dispersion relations and characterization of 
excitation modes both in single component BEC \cite{ticknor_11} and TBEC 
\cite{ticknor_14} have been investigated in the miscible and immiscible 
(side by side configuration) phases. In the present study, we study dispersion 
relations in the immiscible phase for both the side by side and shell 
structured density configurations. More important, the effect of radial 
anisotropy on dispersion relations are investigated for both the 
configurations. A recent work \cite{klaiman_17} has also reported the pathway
from condensation towards fragmentation arising from the anisotropy of 
the confining potential. We also observe anisotropy enhanced fragmentation of 
the outer species in a shell structured immiscible TBEC. 

The purpose of this paper is to present a
systematic study to capture the influences of radial anisotropy in segregated
condensate mixtures. This paper 
is organized as follows. In Sec.~\ref{theory} we provide a brief description
of the HFB-Popov formalism for quasi-2D BEC. We then give the
discrete dispersion relation used in the numerical computations.
The results and discussions are presented in Sec.~\ref{results} for two
representative TBECs, namely $^{85}$Rb~-$^{87}$Rb and $^{133}$Cs~-$^{87}$Rb
TBECs. Sec.~\ref{Rb} presents the discussions on the effects of radial 
anisotropy on density profiles and mode evolution of $^{85}$Rb~-$^{87}$Rb 
TBEC in immiscible regime. This is followed with computation of dispersion 
relation and the consequences of radial anisotropy on it
presented in Sec.~\ref{rb-disp}. In the next two sections, Sec.~\ref{CsRb}
and Sec.~\ref{cs-disp},  we discuss the density profile, mode evolution
and dispersion relation of the $^{133}$Cs~-$^{87}$Rb TBEC. We, then, end
the main part of the paper with conclusions in Sec.~\ref{conc}.


\section{Theory}
\label{theory}
Mean field calculations based on Popov approximation to Hartree-Fock
Bogoliubov theory (HFB) has been of paramount importance in determining the
finite temperature effects and frequencies of collective excitations. To 
start with, we briefly describe the HFB theory for a quasi-2D (pancake shaped) 
TBEC trapped in an anisotropic harmonic potential. This implies that
the frequencies of the harmonic potential satisfies the conditions 
$\omega_x, \omega_y  \ll\omega_z$ and $\omega_x \neq \omega_y$. So, the 
confining potential is of the form
\begin{equation}
V_k(x,y,z)
=(1/2)m_{k}\omega_{x}^2 (x^2+\alpha^2 y^2 + \lambda^2z^2),
\end{equation}
where, $\alpha = \omega_y/\omega_x$, and $\lambda = \omega_z/\omega_x$ are the
anisotropy parameters. In terms of these parameters the requirement to 
have a quasi-2D geometry is $\lambda\gg 1$, $\hbar\omega_z\gg 
\mu_k$. This strongly confines the 
motion of the trapped atoms along the $z$-axis and in this direction the
atoms are frozen in the ground state \cite{petrov_2000}. However, 
excitations are allowed along the $xy$-plane and 
making the system kinematically 2D. Under the mean field approximation, a 
quasi-2D TBEC of interacting bosons is described by the grand-canonical 
Hamiltonian 
\begin{eqnarray}
\hat{H}&=&\sum_{\substack{k=1,2}}\iint dxdy\hat{\Psi}_k^{\dagger}(x,y,t)
\Bigg[-\frac{\hbar^2}{2m_k}\nabla^2_{\perp}+V_k(x,y)-\mu_k 
                       \nonumber \\
       && +\frac{U_{kk}}{2}\hat{\Psi}_k^{\dagger}(x,y,t)\hat{\Psi}_k(x,y,t)
\Bigg]
\hat{\Psi}(x,y,t) +~~U_{12}\iint dxdy
                       \nonumber \\
       && \times \hat{\Psi}_1^{\dagger}(x,y,t)
\hat{\Psi}_2^{\dagger}(x,y,t)\hat{\Psi}_1(x,y,t)\hat{\Psi}_2(x,y,t),
\label{ham}
\end{eqnarray}
with $k=1,2$ denoting the species index, 
$\hat{\Psi}_k$ ($\hat{\Psi}_k^{\dagger}$) are the Bose field 
annihilation (creation) operators of the two species, and $\mu_k$s are the 
chemical potentials. In quasi-2D, $U_{kk} = 2a_{kk}\sqrt{2\pi\lambda}$ and
$U_{12}=2a_{12}\sqrt{2\pi\lambda}(1+ m_1/m_2)$ are the intra-species and 
inter-species interactions, respectively. Here, $a_{kk}$, $a_{12}$ are
the $s$-wave intra- and interspecies scattering lengths. In the present work
we consider only repulsive interactions and hence, $a_{kk}>0$, $a_{12}>0$. 
From the Hamiltonian in Eq.~(\ref{ham}), the dynamics of Bose field operators 
are given by the coupled equations
\begin{eqnarray}
\begin{aligned}
 i\hbar
 \begin{pmatrix}
   \dot{\hat{\Psi}}_1\\
   \dot{\hat{\Psi}}_2
\end{pmatrix} \!\!
= \!\!\begin{pmatrix}
  \hat{h}_1 + U_{11}\hat{\Psi}_1^\dagger\hat{\Psi}_1 & U_{12}
           \hat{\Psi}_2^\dagger \hat{\Psi}_1\\
   U_{12}\hat{\Psi}_1^\dagger\hat{\Psi}_2             & \hat{h}_2
           + U_{22}\hat{\Psi}_2^\dagger \hat{\Psi}_2
  \end{pmatrix} \!\!\!
  \begin{pmatrix}
    \hat{\Psi}_1\\
    \hat{\Psi}_2  
  \end{pmatrix}, 
\label{twocomp}
\end{aligned}
\end{eqnarray}
where, $\hat{h}_{k}= (-\hbar^{2}/2m_k)\nabla_\perp^2 +V_k(x,y)-\mu_k$ is the
single-particle part of the Hamiltonian. 

When the temperature is below the critical temperature ($T_c$), 
majority of the atoms occupy the ground state to form a condensate. Thus for 
$T\ll T_c$ following the Bogoliubov decomposition, the Bose field operator can 
be expressed as the sum of condensate part and the fluctuations over it 
$\hat{\Psi}_k = \phi_k + \tilde{\psi}_k$, where
$\phi_k=\langle\hat{\Psi}_k\rangle$s are the $c$-fields representing each
of the condensate species, and $\tilde{\psi}_k$s are the corresponding
non-condensate densities or fluctuations which may be either quantum or 
thermal in nature. By definition the fluctuation operators satisfy the 
condition
$\langle\tilde{\psi}_k\rangle = \langle\tilde{\psi}_k^\dagger\rangle = 0$.
In addition, on the application of time-independent HFB-Popov approximation 
\cite{griffin_96},  Eq.~(\ref{twocomp}) reduces to the coupled generalized 
Gross-Pitaevskii (CGGP) equations 
\begin{eqnarray}
 \hat{h}_k\phi_k + U_{kk}\left[n_{ck}+2\tilde{n}_{k}\right]\phi_k
  +U_{12}n_{3-k}\phi_k=0,
\label{gpem}
\end{eqnarray}
where, $\phi_k$s is the stationary solutions of CGGP with 
$n_{ck}(x,y) \equiv |\phi_k(x,y)|^2$, $\tilde{n}_k(x,y) 
\equiv \langle\tilde{\psi}^{\dagger}(x,y,t)\tilde{\psi}(x,y,t)\rangle$
and $n_k(x,y)=n_{ck}(x,y)+\tilde{n}_k(x,y)$ represent the local
condensate, non-condensate and total density respectively. Employing Bogoliubov
transformation, fluctuations $\tilde{\psi}_k$ and its complex conjugate 
$\tilde{\psi}_k^{\dagger}$ are expressed as the linear combination of
quasiparticle creation ($\hat{\alpha}_j^{\dagger}$) and annihilation 
($\hat{\alpha}_j$) operators \cite{ticknor_14}
\begin{eqnarray}\label{bgtrans}
\begin{aligned}
\tilde{\psi}_k &=&\sum_{j}\big[u_{kj}(x,y)\hat{\alpha}_j
e^{-i\varepsilon_{j}t/\hbar}- v_{kj}^{\ast}(x,y)
\hat{\alpha}_j^{\dagger}e^{i\varepsilon_{j}^{\ast}t/\hbar}\big],\\
\tilde{\psi}_k^{\dagger} &=& \sum_{j}\big[u_{kj}^{\ast}(x,y)
\hat{\alpha}_j^{\dagger}
e^{i\varepsilon_{j}^{\ast}t/\hbar}- v_{kj}(x,y)
\hat{\alpha}_je^{-i\varepsilon_{j}t/\hbar}\big],
\end{aligned}
\end{eqnarray}
where $j$ is the index representing the sequence of the quasiparticle
excitation. The quasiparticle creation and annihilation operators satisfy the 
usual Bose commutation relations. $u_{jk}$ and $v_{jk}$ are complex functions 
and denote the Bogoliubov quasiparticle amplitudes with the normalisation
\begin{equation}\label{norm}
\iint dxdy\sum_k\left(|u_k(x,y)|^2-|v_k(x,y)|^2\right)=1.
\end{equation}
Considering all the above decompositions and transformations, the equation of
motion of the fluctuation operator are 
\begin{subequations}
\begin{eqnarray}
 \hat{{\mathcal L}}_{1}u_{1j}-U_{11}\phi_{1}^{2}v_{1j}+U_{12}\phi_1 \left 
   (\phi_2^{*}u_{2j} -\phi_2v_{2j}\right )&=& \varepsilon_{j}u_{1j},
\;\;\;\;\;\;\\
    \hat{\underline{\mathcal L}}_{1}v_{1j}+U_{11}\phi_{1}^{*2}u_{1j}-U_{12}
    \phi_1^*\left (\phi_2v_{2j}-\phi_2^*u_{2j} \right ) 
     &=& \varepsilon_{j}v_{1j},\;\;\;\;\;\;\\
    \hat{{\mathcal L}}_{2}u_{2j}-U_{22}\phi_{2}^{2}v_{2j}+U_{12}\phi_2\left 
    ( \phi_1^*u_{1j}-\phi_1v_{1j} \right ) &=& \varepsilon_{j}u_{2j},
\;\;\;\;\;\;\\
\hat{\underline{\mathcal L}}_{2}v_{2j}+U_{22}\phi_{2}^{*2}u_{2j}-U_{12} 
\phi_2^*\left ( \phi_1v_{1j}-\phi_1^*u_{1j}\right ) &=& 
\varepsilon_{j}v_{2j},\;\;\;\;\;\;\;\;\;
\end{eqnarray}
\label{bdg2m}
\end{subequations}
which are referred as the Bogoliubov-de-Gennes equations for the quasi-2D TBEC
system \cite{ticknor_13,roy_14a} with $\hat{{\mathcal L}}_{1}=
\big(\hat{h}_1+2U_{11}n_{1}+U_{12}n_{2})$, 
$\hat{{\mathcal L}}_{2}=\big(\hat{h}_2+2U_{22}n_{2}+U_{12}n_{1}\big)$ 
and $\hat{\underline{\cal L}}_k  = -\hat{\cal L}_k$. 

To solve Eq.~(\ref{bdg2m}), $u_{kj}$s and $v_{kj}$s are written as a linear 
combination of the harmonic oscillator eigenstates and the 
Bogoliubov-de-Gennes matrix (BdGM) constructed from Eq.~(\ref{bdg2m}) is 
diagonalized. Eq.~(\ref{bdg2m}) along with Eq.~(\ref{gpem}) are known as 
Hartree-Fock Bogoliubov (HFB) equations and need to be solved self 
consistently. The solutions are the order parameters $\phi_k$s and the 
non-condensate densities $\tilde{n}_k$s. The thermal components, in terms of 
the quasiparticle amplitudes, are 
\begin{equation}
 \tilde{n}_k=\sum_j\left [(|u_{kj}|^2+|v_{kj}|^2)N_{0}(\varepsilon_{j})+
|v_{kj}|^2\right ],
 \label{n_k2}
\end{equation}
where, $N_0(\varepsilon_{j}) = (e^{\beta \varepsilon_{j}} - 1) ^{-1}$ with 
$\beta=1/(k_{\rm B}T) $ is the Bose factor of the $j$th quasiparticle mode at 
temperature $T$. As $T$ approaches to zero, the role of thermal fluctuations 
gradually diminishes. At $T = 0$, thermal fluctuations is absent and the
non-condensate density arises only from the quantum fluctuations, which is 
evident from Eq.~(\ref{n_k2}) as it is reduced 
to $\tilde{n}_k=\sum_j|v_{kj}|^2$ at $T = 0$. Our studies show that the
$\tilde{n}_k$ arising from the quantum fluctuations does not produce
significant changes in the BdG spectrum, and so, we avoid the condition of 
self-consistency at zero temperature. The details of the numerical
scheme adopted to compute the Bogoliubov quasiparticle amplitudes and 
fluctuations has been elaborated in Ref.~\cite{pal_17}.

The dispersion relations determine the response of a physical system when 
subjected to external perturbations. For the present work,
the change in the geometry of the interface in a TBEC arising from a change in
the anisotropy of the confining potential affects the energies and amplitudes 
of the quasiparticle excitations. In the limit of low $\alpha$, the TBECs 
consist of three topologically connected fragments. The dispersion relation is 
thus expected to change. To define the dispersion relation we compute the 
root mean square of the wave number $k^{\rm rms}$ of each quasiparticle mode,
and the results  describes a discrete dispersion relation. Following 
Refs.~\cite{ticknor_14, wilson_10}, the $k^{\rm rms}$ of the $j$th 
quasiparticle is
\begin{equation}
 k_j^{\rm rms} = \left\{\frac{\sum_i\int d\mathbf{k} k^2
                 [|u_{ij}(\mathbf{k})|^2 + |v_{ij}(\mathbf{k})|^2]}
                 {\sum_i\int d\mathbf{k} [|u_{ij}(\mathbf{k})|^2 
                 + |v_{ij}(\mathbf{k})|^2]}\right\}^{1/2}.
  \label{dspeq}
\end{equation}
It is to be noted here that $k_j^{\rm rms}$ are defined in terms of the
quasiparticle modes corresponding to each of the constituent species
defined in the $k$ or momentum space through the index $i = 1, 2$. It is 
then essential to compute $u_j(\mathbf{k})$ and $v_j(\mathbf{k})$, the 
Fourier transform of the Bogoliubov quasiparticle amplitudes $u_j(x,y)$ 
and $v_j(x,y)$, respectively. Once we have $k_j^{\rm rms}$ for all
the modes we obtain a dispersion curve, and we can then examine how the
change in the condensate topology modifies the dispersion curve.

\section{Results and Discussions}
\label{results}

In a trapped quasi-2D TBEC, the low-lying BdG spectrum supports two Goldstone
modes for each of the condensate species due to breaking of
$U(1)$ global gauge symmetry. The two lowest modes corresponding to each of the
species with non zero excitation energies are called the dipole 
modes. The dipole modes which oscillate out-of-phase with each other are called slosh modes. In-phase slosh modes with center-of-mass motion are called 
the Kohn modes. 
The frequency of the Kohn mode is independent of the type of interactions and 
interaction strength as well. The presence of the Kohn modes in the trapped 
system follows from Kohn's theorem \cite{kohn_61, fetter_98}. According to the 
theorem BEC in a confining potential must have a mode in which the centre of 
mass oscillates with the frequency of the confining potential. However, in a 
magnetic trap in the vicinity of the Feshbach resonance \cite{jibbouri_13} and 
in few-electron parabolic quantum dots doped with a single magnetic 
impurity \cite{nguyen_10}, there are deviations from Kohn's theorem.

 A variation in the interaction strength drives the TBEC system from 
immiscible to miscible configuration or vice versa. In addition, it can induce
dynamical instability resulting in the swapping of the species. But, the radial
anisotropy can alter the equilibrium density distribution of the ground state 
and modify the evolution of low lying mode energies. Among the low lying modes 
of the BdG spectrum, the dipole (Kohn/slosh) and quadrupole modes have 
dominant contribution  to $\tilde{n}$ at $T\sim 0$K and hence we examine these
modes detail. The transformation of these modes due to radial anisotropy
is one of the primary issues addressed in this paper, and as example we 
consider specific systems.
\begin{figure}[H]
 \includegraphics[width=8.5cm]{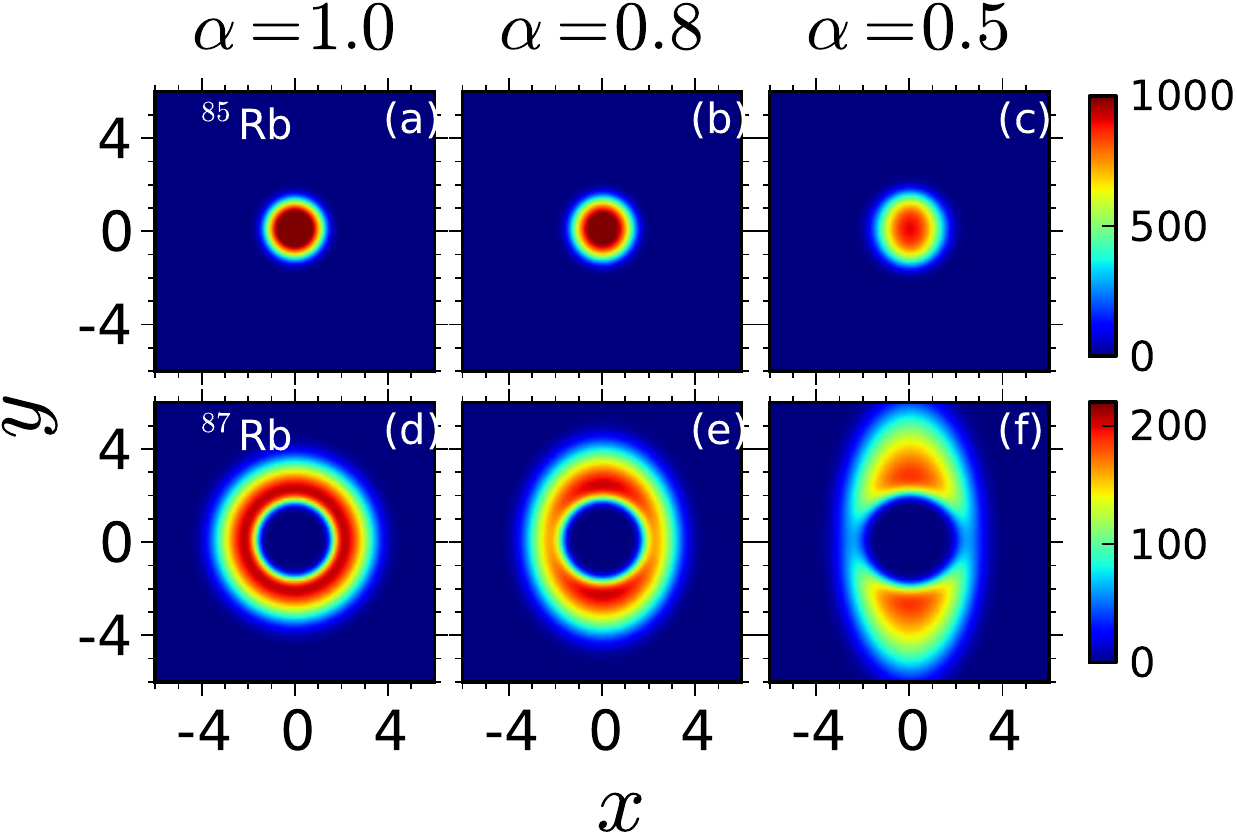}
    \caption{(Color online) Equilibrium condensate density profiles of 
             $^{85}$Rb~-$^{87}$Rb TBEC at zero temperature for three different
             values of $\alpha$ ($\alpha$ = 1.0, 0.8, 0.5). 
             $a_{11}$ is kept fixed at $10a_0$ . (a)-(c) Equilibrium density
             profiles corresponding to $^{85}$Rb. (d)-(f)Equilibrium density
             profiles corresponding to $^{87}$Rb. $n_c$ is measured in units of
             $a_{\rm osc}^{-2}$ and the spatial coordinates $x$ and $y$ are
             measured in units of $a_{\rm osc}$.}
    \label{den10}
\end{figure}


\subsection{Mode evolution of $^{85}$Rb~-$^{87}$Rb BEC mixture}
\label{Rb}

In this section, we consider the TBEC of homonuclear atoms at zero temperature 
with shell structured density profile in immiscible domain. We choose 
$^{85}$Rb~-$^{87}$Rb TBEC as a representative example. For convenience, we 
designate  atoms of $^{85}$Rb and $^{87}$Rb as species 1 and 2, respectively. 
The intra-species scattering length of $^{87}$Rb atoms has the value
$a_{22} = 99a_0$, and the inter-species scattering length 
$a_{12} = 214a_0$. We consider equal number of atoms for the two species, 
i.e,  $N_1 = N_2 = 5000$ and the quasi-2D Rb-Rb TBEC that we considered here
is obtained with trapping parameters $\lambda = 12.5$ and 
$\omega_x = 2\pi \times 8.0$Hz \cite{neely_10}. Here, it is to be emphasized
that the background intra-species scattering length of $^{85}$Rb ($a_{11}$)
is negative, and to obtain BEC of $^{85}$Rb it is essential to tune the 
scattering length to positive values using magnetic Feshbach resonance 
\cite{cornish_00}. For the present work we perform our calculations for three
different values of $a_{11} = 10a_0, 50a_0, 180a_0$. 
With this specific set of trapping and interaction parameters, at equilibrium, 
the ground state of the TBEC is phase-separated with shell structured density 
profile. For $\alpha=1$, (i.e., for $\omega_x=\omega_y$) the TBEC is 
rotationally symmetric and the interface separating the two condensates is 
circular. However, for $\alpha<1$ the rotational symmetry is broken with
the density profile of the condensates elongated along $y$ direction and as 
$\alpha$ is decreased the interface transforms from circular to planar
\begin{figure}[H]
 \includegraphics[width=8.5cm]{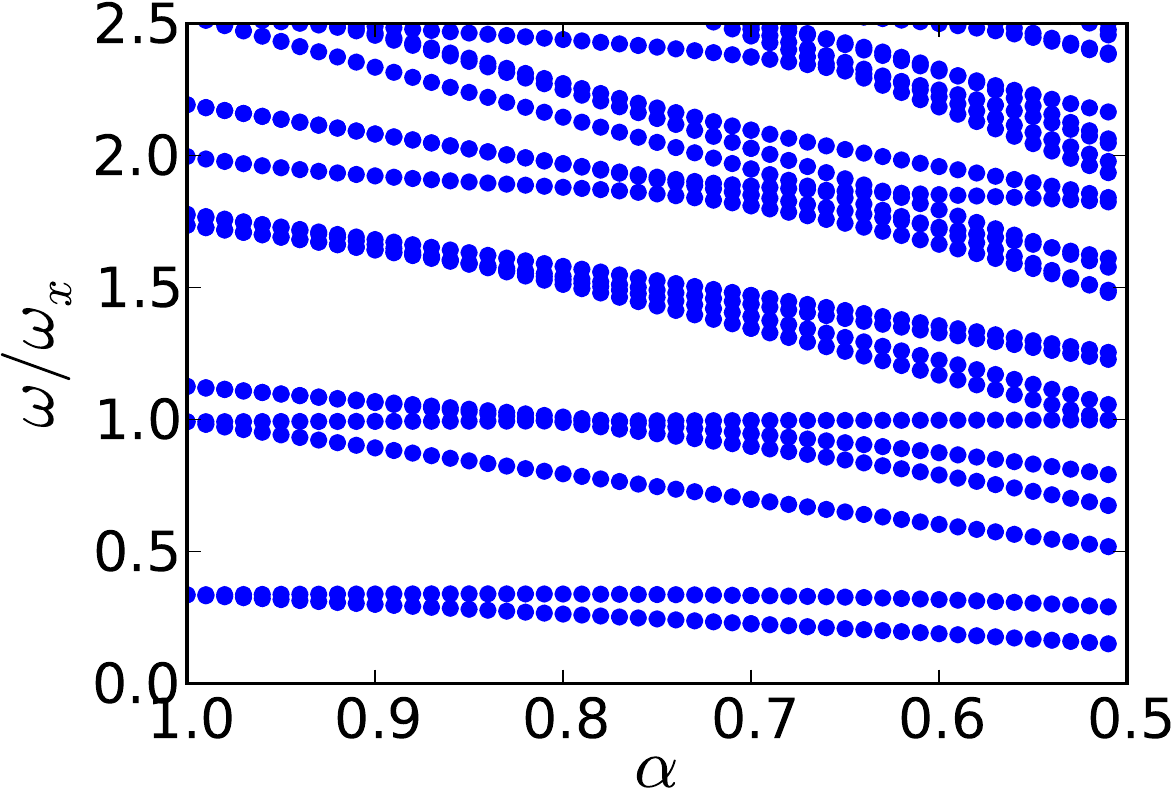}
    \caption{(Color online) The evolution of low-lying quasiparticle modes in
             the phase separated domain of $^{85}$Rb~-$^{87}$Rb TBEC
             as a function of $\alpha$ in the domain $0.5 \le \alpha \le 1.0$
             for $a_{11} = 10a_0$ at zero temperature with $N_1 = N_2 = 5000$.}
 \label{mevlRbRb10}
\end{figure}

In Fig.~\ref{den10} we show the equilibrium condensate density profiles for 
$a_{11}$ = 10 $a_0$ for $\alpha=1.0, 0.8$ and $0.5$. As the profiles show
TBEC is phase-separated with $^{85}$Rb atoms lying at the center surrounded 
by the $^{87}$Rb atoms. With decreasing $\alpha$, the density profiles are
modified and are fragmented at lower values of $\alpha$. In the figure, the
fragmentation is discernible for $\alpha = 0.5$, where the condensate density 
profile of $^{87}$Rb appears to consist of two symmetric lobes equidistant 
from the origin. For all the three cases, $^{85}$Rb condensate remains 
at the core and with lower values of $\alpha$ it expands along $y$-direction 
and changes its geometry from circular to ellipse with $y$-axis being the major
axis. The change in the density distribution also affects the quasiparticle
excitation spectrum. For $a_{11} = 10a_0$, the quasi particle mode evolution
as a function of $\alpha$ is shown in Fig.~\ref{mevlRbRb10}. As seen from the
figure the degeneracies of the mode energies are lifted when  $\alpha <1$.
For the dipole mode, as the degeneracy is lifted the Kohn mode
remains steady at $\omega = \omega_x$ whereas the energy of the slosh mode
decreases. 
\begin{figure}[H]
 \includegraphics[width=8.5cm]{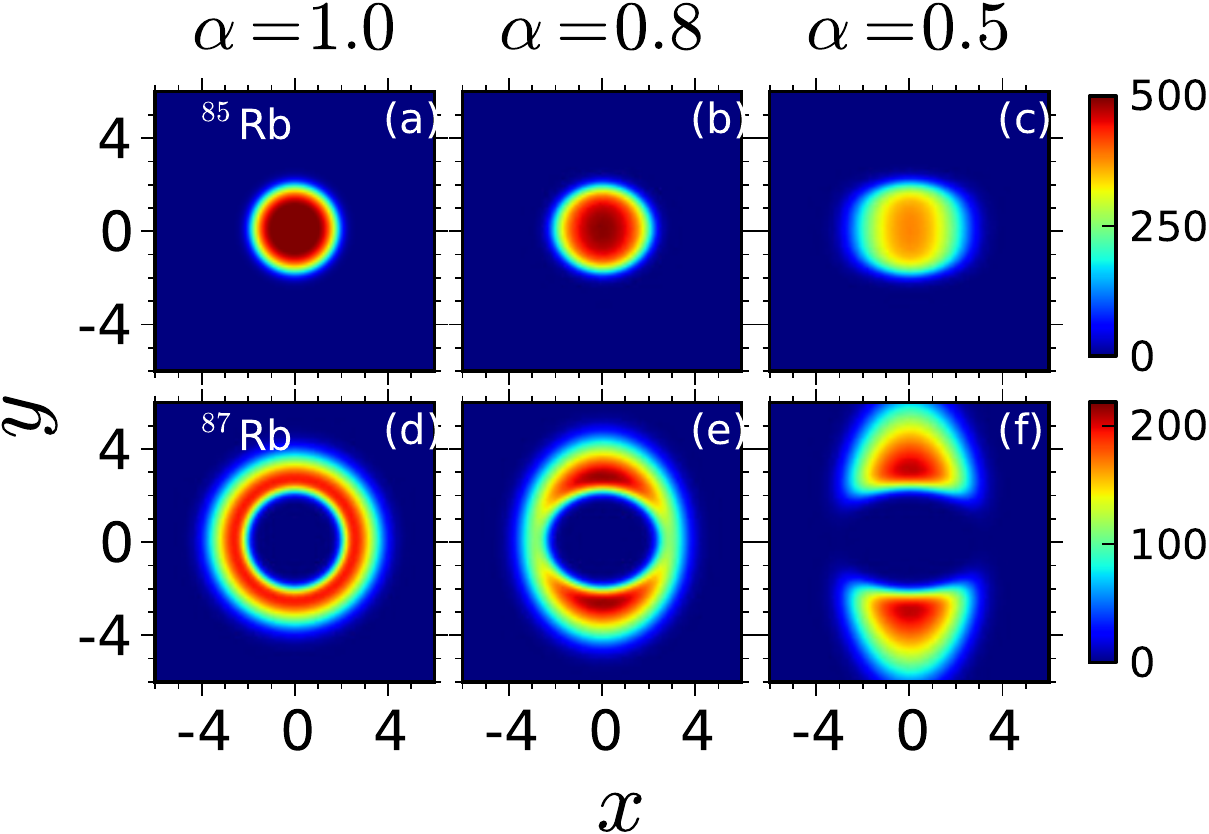}
    \caption{(Color online) Equilibrium condensate density profiles of 
             $^{85}$Rb~-$^{87}$Rb TBEC at zero temperature for three different
             values of $\alpha$ ($\alpha$ = 1.0, 0.8, 0.5). $a_{11}$ remains
             fixed at $50a_0$ . (a)-(c) Equilibrium density profiles
             corresponding to $^{85}$Rb  (d)-(f)Equilibrium density profiles
             corresponding to $^{87}$Rb. Equilibrium condensate density $n_c$
             is measured in units of $a_{\rm osc}^{-2}$ and the spatial
             coordinates $x$ and $y$ are measured in units of $a_{\rm osc}$.}
 \label{den50}
\end{figure}

\begin{figure}[H]
 \includegraphics[width=8.5cm]{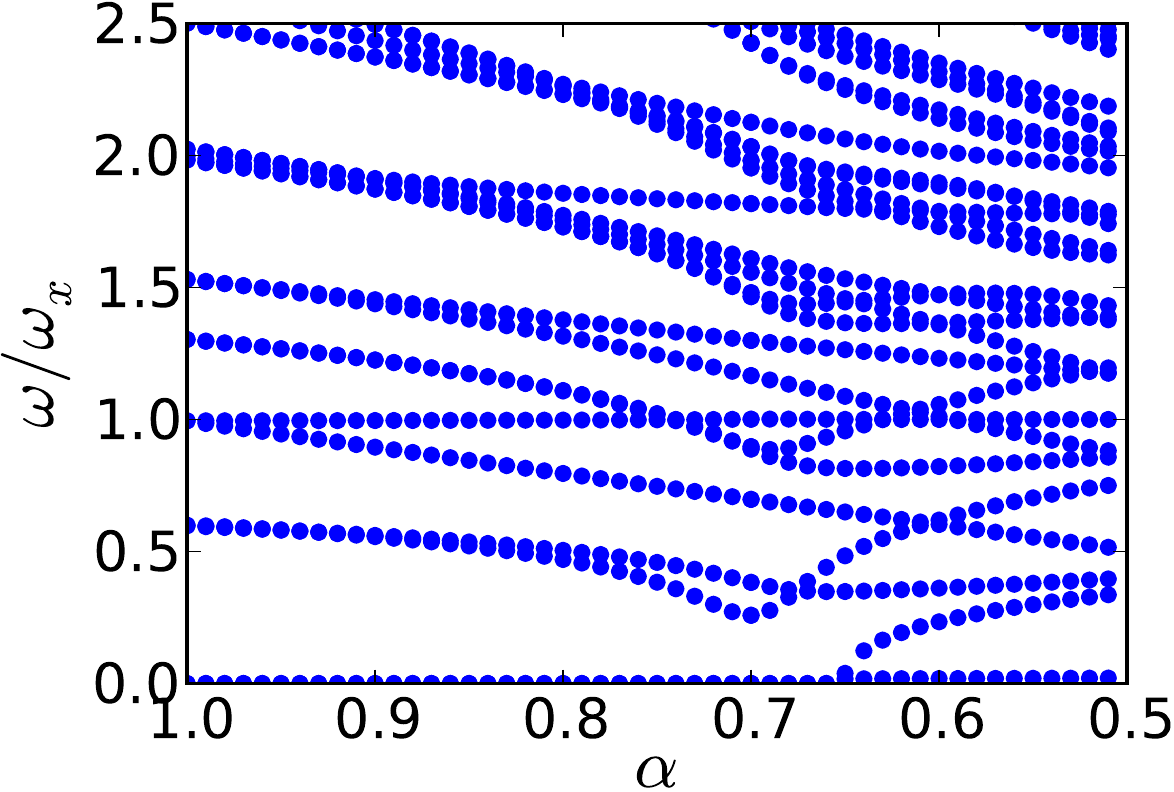}
    \caption{(Color online) The evolution of low-lying mode energies as a
             function of $\alpha$ in the phase separated domain of 
             $^{85}$Rb~-$^{87}$Rb TBEC at zero temperature. $\alpha$ is
             varied in the domain $0.5 \leqslant \alpha \leqslant 1.0$ 
             while $a_{11}$ is 
             kept fixed at $50a_0$. For $^{85}$Rb~-$^{87}$Rb TBEC, it is a
             domain of interest since dynamical instability sets in at 
             $a_{11} = 50a_0$   for $N_1 = N_2 = 5000$.}
 \label{mevlRbRb50}
\end{figure}

 For the isotropic case ( $\alpha = 1$) the $^{85}$Rb~-$^{87}$Rb TBEC is
dynamically unstable in the region $50a_0 \leqslant a_{11} \leqslant 150a_0$ 
\cite{pal_17}. The presence of anisotropy affects the dynamical instability of the mixture and with the increase of $\alpha$ instability decreases. Thus, both the onset of instability and the radial anisotropy
can have a major influence on the equilibrium density profile at 
$a_{11} = 50a_0$. As shown in Fig.~\ref{den50}, the density profiles do 
exhibit a transformation as a function of $\alpha$ at $a_{11} = 50a_0$. The 
equilibrium density profile in Fig.~\ref{den50}(f) shows that unlike 
Fig.~\ref{den10}, the two lobes of $^{87}$Rb are disconnected. In addition, 
when $a_{11} > a_{22}$, the species swaps their positions and the density
profile of $^{85}$Rb~-$^{87}$Rb TBEC has similar configuration as was for 
$a_{11} < a_{22}$. For example, with $a_{11}=180a_0$ and $a_{22} = 99a_0$,
the equilibrium density profile of $^{85}$Rb~-$^{87}$Rb TBEC has the similar
configuration as shown in Fig.~\ref{den10} but with species positions
interchanged. To examine the further, the low lying mode energies are shown as 
a function of $\alpha$ in Fig.~\ref{mevlRbRb50}. Unlike in the case of 
$a_{11} = 10a_0$ there is a doubly degenerate zero energy mode. In addition, 
some of the low-lying modes show a decrease in energy up to 
$\alpha \approx 0.7$, and increase for $\alpha < 0.7$. The important point to 
note is that the new zero energy mode remains degenerate upto $\alpha = 0.65$. 
But, for $\alpha < 0.65$ the degeneracy is lifted and the mode energy 
bifurcates into two branches. One branch continues to be the zero energy mode, 
where as the other branch becomes hard. Similarly, the doubly degenerate 
quadrupole mode with energy $\hbar\omega \approx 0.6\hbar\omega_x$ at 
$\alpha = 1$ continue to soften and at $\alpha \approx 0.8$ the degeneracy is 
lifted creating two branches. For $\alpha < 0.8$, both the branches continue 
to soften but at $\alpha \approx 0.7$ the branches start to harden. The hardening and softening of the low energy quasiparticle modes have 
also been observed around the point of phase separation of TBECs in optical
lattices \cite{suthar_15, suthar_16}, and due to the change in the geometry 
and topology of the confining potential \cite{roy_16}. 

\subsection{Excitations and dispersion relations for $^{85}$Rb~-$^{87}$Rb TBEC}
\label{rb-disp}

\begin{figure}[H]
 \includegraphics[width=8.5cm]{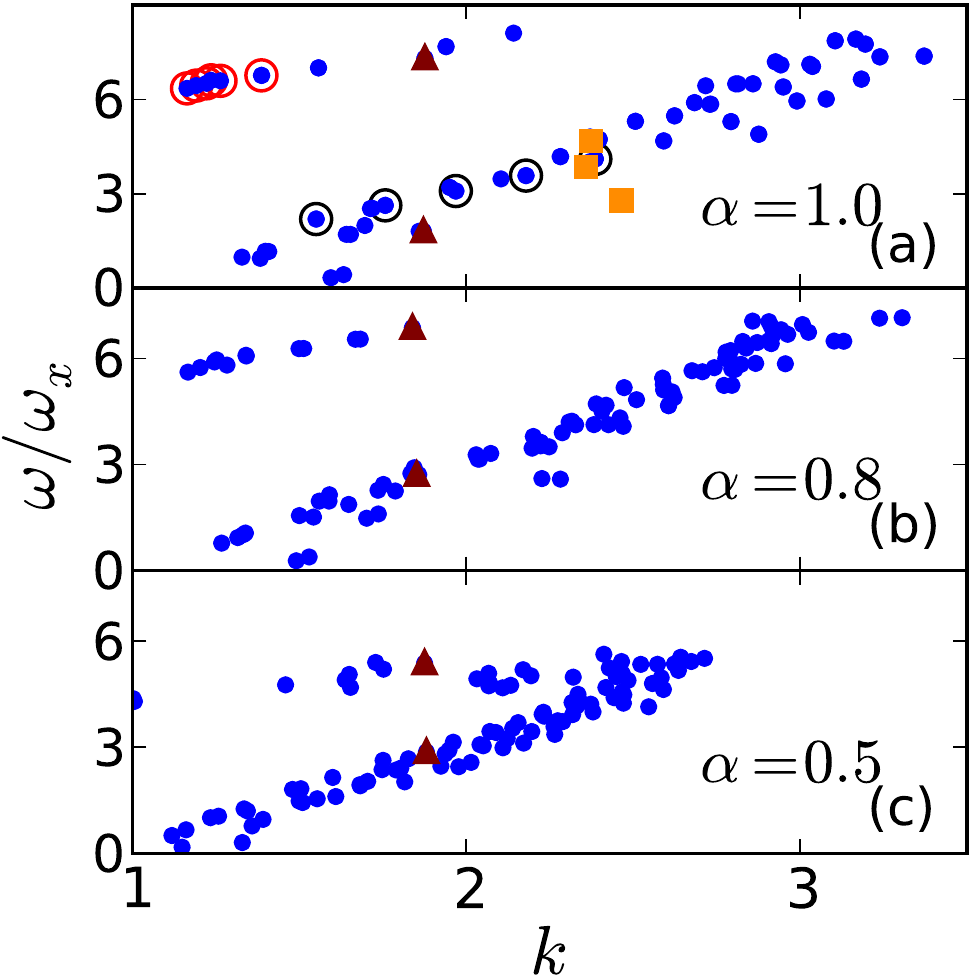}
    \caption{(Color online) (a) The dispersion relation for immiscible 
             (shell structured) $^{85}$Rb~-$^{87}$Rb TBEC at $a_{11}$=10$a_0$
             when $\alpha = 1.0$. In this plot the dispersion curves can be
             broadly classified into two broadly separated branches. The quasi
             particle amplitudes corresponding to the upper branch marked with
             red empty circles $({\color{red}\bigcirc})$ are shown in 
             Fig.~\ref{qamp1-disprb}. For the lower branch the quasi particle
             amplitudes is shown in Fig.~\ref{qamp2-disprb} along a particular
             path marked with empty black circle $({\color{black}\bigcirc})$.
             (b) The dispersion relation for immiscible (shell structured) 
             $^{85}$Rb~-$^{87}$Rb TBEC at $a_{11}$=10$a_0$ when $\alpha = 0.8$.
             (c) The dispersion relation for immiscible (shell structured)
             $^{85}$Rb~-$^{87}$Rb TBEC at $a_{11}$=10$a_0$ when $\alpha = 0.5$.
             As value of $\alpha$ is reduced from 1.0 i.e., radial anisotropy
             makes this two branches comparatively closer to each other.}
 \label{disp-rbrb}
\end{figure}

 The dispersion curves are computed based on Eq.~(\ref{dspeq}). This involves
computing the $k^{\rm rms}$ of the $j$th quasiparticle, and then the 
corresponding quasiparticle mode energies are plotted as function of the
$k^{\rm rms}$. In the miscible domain the dispersion curve is expected to be
devoid of structures as the quasiparticle amplitudes of the component BECs 
are similar. However, in the immiscible domain the vastly different density
distributions of the two species lead to diverse quasiparticle amplitude
structures. And, this can manifest as structures in the dispersion curves.
As a specific example, consider the case of shell structured immiscible 
density distribution of $^{85}$Rb~-$^{87}$Rb TBEC. In spite
of rotational symmetry, due to immiscible domain the dispersion curves shown
in Fig.~\ref{disp-rbrb} exhibit complex trends. 

\begin{figure}[t]
\centering
\includegraphics[width=8.5cm]{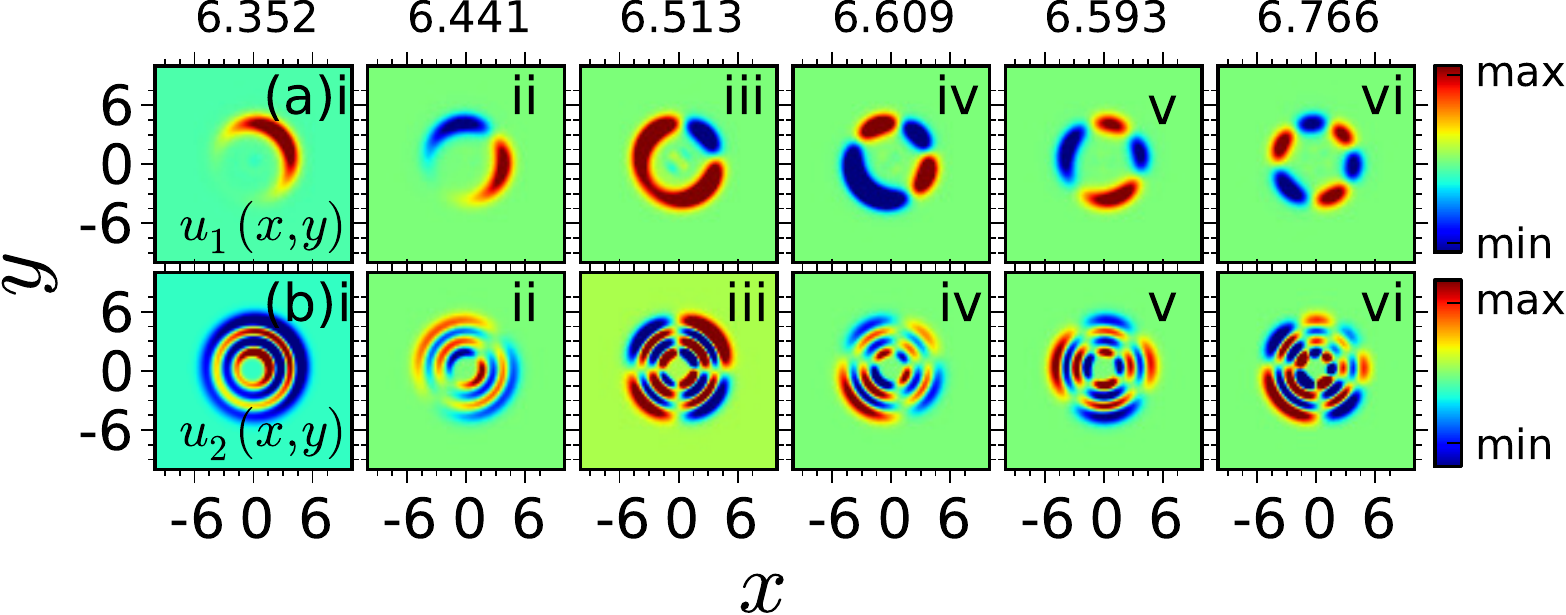}
    \caption{(Color online) Quasiparticle amplitudes corresponding to
             interface excitations for the immiscible 
             (shell structured) $^{85}$Rb~-$^{87}$Rb TBEC in space at 
             $\alpha =1.0$. (a)i - vi: Quasi particle amplitudes for species 
             $^{85}$Rb. (b)i - vi: Quasiparticle amplitudes for
             species $^{87}$Rb. Each set of
             quasiparticles start from small values of $k$ (at left) and move
             towards large values of $k$ at right. In dispersion curve given
             in Fig.~\ref{disp-rbrb}(a), these points are marked with empty red
             circles $({\color{red}\bigcirc})$. Excitation energy for each set
             of $u_1(x,y)$ and 
             $v_1(x,y)$ are highlighted at the top of the $u_1(x,y)$. 
             Here $u_1$s and $u_2$s are in units of 
             $a_{\rm osc}^{-1}$. Spatial coordinates $x$ and $y$ are measured in
             units of $a_{\rm osc}$.}
 \label{qamp1-disprb}
\end{figure}

  In Fig.~\ref{disp-rbrb}(a) the dispersion relation at $\alpha = 1.0$ for 
$^{85}$Rb~-$^{87}$Rb TBEC in immiscible (sandwich) domain with 
$a_{11} = 10a_0$ is shown. The dispersion curve has two well separated 
branches: the {\em upper branch} starts at $\approx6.35\hbar\omega_x$ and the 
{\em lower branch} starts close to zero energy. This is in contrast to the 
dispersion relations reported for the immiscible TBEC with side by side 
configuration. One important reason to study dispersion curves is to identify 
modes with different characteristics. In the present case the two well 
separated modes correspond to two different classes of excitations: the 
interface and bulk excitations. To analyze these two branches in better detail 
we plot selected quasiparticle amplitudes from these two branches. For the 
lower branch, the quasiparticle amplitudes of the 
modes marked with red empty circle $({\color{red}\bigcirc})$ are shown in 
Fig.~\ref{qamp1-disprb}. These modes with low-$k$ have high energies and are
localized at the interface of the two species. 

  In Fig.~\ref{qamp1-disprb}(a)(i-vi) the interface excitations for $^{85}$Rb,
the condensate species which occupies the central region, are shown. The 
interface nature of the quasiparticles is discernible from the geometry as 
these have non-zero values only at the interface. With the increase of energy,
numerical values annotated at the top of each column, azimuthal quantum 
number ($m$) increases. It must be mentioned that in the figures, for 
information the energies are listed up to the fifth decimal, and this is not
an indication of the accuracy. So, here after for the description of the 
results we list only up to the second decimal. To illustrate the change in
$m$ consider Fig.~\ref{qamp1-disprb}(a)(ii), which corresponds to mode with
energy $6.44\hbar\omega_x$ and corresponds to $m = 1$. Whereas the mode
in Fig.~\ref{qamp1-disprb} (a)(vi) which has larger number of lobes and 
corresponds to $m = 3$ has energy $6.77\hbar\omega_x$. On the other hand, 
Fig.~\ref{qamp1-disprb}(b)(i-vi) show the interface excitations for 
$^{87}$Rb. Since it is energetically favorable for the outer species to expand,
the quasiparticles corresponding to $^{87}$Rb are composed of four radial 
nodes with $m$ being directly proportional to the mode energy.

\begin{figure}[H]
\centering
 \includegraphics[width=9.0cm]{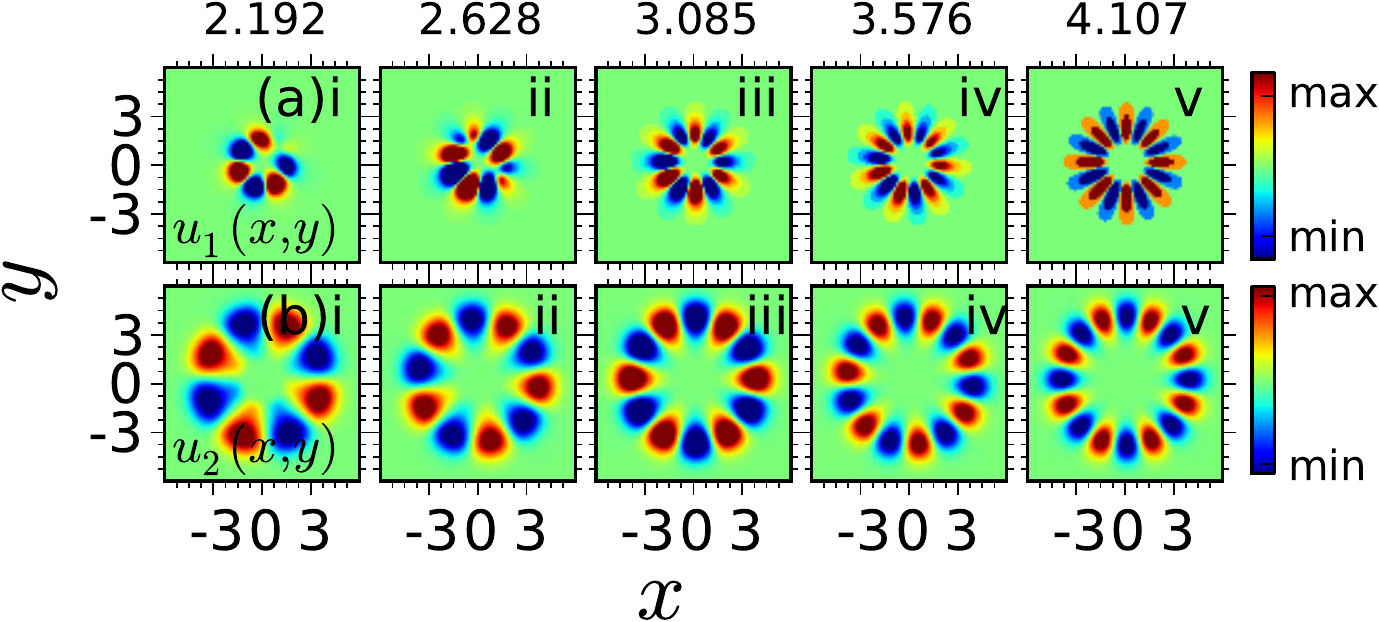}
    \caption{(Color online) Quasiparticle amplitudes corresponding to bulk
             excitations for the immiscible 
             (shell structured) $^{85}$Rb~-$^{87}$Rb TBEC in space for 
             $\alpha = 1.0$. (a)i - v: Quasi particle amplitudes for species
             $^{85}$Rb and (b)i - v: Quasiparticle
             amplitudes for species $^{87}$Rb. Along this branch both the
             species have one radial nodes but the azimuthal quantum number
             increases with the increase of mode energy. Each set of
             quasiparticles starts from small values of
             $k$ (at left) and move towards large values of $k$ at right.
             In dispersion curve given in Fig.~\ref{disp-rbrb}(a), these
             points are marked with empty black circles
             $({\color{black}\bigcirc})$. 
             Excitation energy for each set of $u_1(x,y)$ and $v_1(x,y)$ are
             highlighted at the top of the $u_1(x,y)$. 
             Here $u_1$s and $v_1$s are in units of $a_{\rm osc}^{-1}$. Spatial
             coordinates $x$ and $y$ are measured in units of $a_{\rm osc}$.}
 \label{qamp2-disprb}
\end{figure}

For the lower branch the quasiparticle amplitudes of few specific modes marked 
with black empty circle $({\color{black}\bigcirc})$ are shown in 
Fig.~\ref{qamp2-disprb}. These are the bulk excitations of the TBEC mixture.
Like in the previous case, Fig.~\ref{qamp2-disprb}(a)(1-6) show the 
quasiparticle amplitudes for $^{85}$Rb, and Fig.~\ref{qamp2-disprb}(b)(1-6) 
correspond to $^{87}$Rb. The modes are marked with black empty circle 
$({\color{black}\bigcirc})$ and these trace a path in the dispersion curve
from small-$k$ towards large-$k$. Along the path the quasiparticles have one
radial node and it is discernible in the quasiparticle amplitudes of both the
species. The azimuthal quantum number $m$ increases with the increase of mode 
energy. The bulk nature of the excitations is manifest from the structure
of the quasiparticle amplitudes as they coincide with the condensate density 
distribution. 

We find that the other modes in the upper and lower branches have similar 
trends as described earlier. However, for clarity we also examine the structure
of the modes marked in with orange square 
$({\color{orange(colorwheel)}\blacksquare})$ and maroon triangle 
$({\color{maroon(html/css)}\blacktriangle})$ in Fig.~\ref{disp-rbrb} given in 
the appendix. The modes considered have higher $k$ but the interface and bulk
nature of the modes are discernible from their spatial structures. To show the 
effect of the radial anisotropy, the dispersion curves for $\alpha = 0.8$ and 
$\alpha = 0.5$ are plotted in Fig.~\ref{disp-rbrb}(b-c). From the plots we can 
observe that radial anisotropy decreases the separation between the two 
branches. Another important impact of higher anisotropy the spacing of the 
modes in $k$ space decreases. That is, the modes with $k$ in the range 
1.16-3.37 is reduced to 1.00-2.71 when the $\alpha = 0.5$, and the two 
branches merge at larger $k$.


\subsection{Mode evolution of $^{133}$Cs~-$^{87}$Rb BEC mixture}
\label{CsRb}
In this section, we consider the TBEC of heteronuclear atoms where the
density profile has side by side configuration in the immiscible domain.
As a specific example we consider  $^{133}$Cs~-$^{87}$Rb TBEC at 
$T = 0$K, however, the results obtained are generic to TBECs of two different
atomic species. In this system, we consider Cs and Rb
to be species 1 and 2, respectively.  With this identification, the $s$-wave
scattering lengths corresponding to intra-species interactions of Cs and Rb
are $a_{\rm Cs} = a_{11} = 280a_0$ and $a_{\rm Rb} = a_{22} = 100a_0$
respectively, where as mentioned earlier $a_0$ denotes the Bohr radius. The
TBEC contains equal number of atoms for the two species 
i.e.,  $N_1 = N_2 = 2000$, and the trapping parameters are the same as
considered for $^{85}$Rb~-$^{87}$Rb TBEC in Sec.~\ref{Rb}. The $s$-wave
scattering length for inter-species interaction is taken as
$a_{\rm CsRb} = a_{12} = 220a_0$, which is less than the background 
inter-species 
scattering length $a_{\rm CsRb} = 650a_0$ \cite{lercher_11}.
With this value of $a_{12}$, at $T = 0$, $^{133}$Cs~-$^{87}$Rb TBEC is in
the immiscible domain (as dictated by the condition of phase separation---
$a_{12}\gg \sqrt{a_{11}a_{22}}$ under Thomas-Fermi limit at $T = 0$ ) with
side by side density configuration. 

\begin{figure}[H]
 \includegraphics[width=8.5cm]{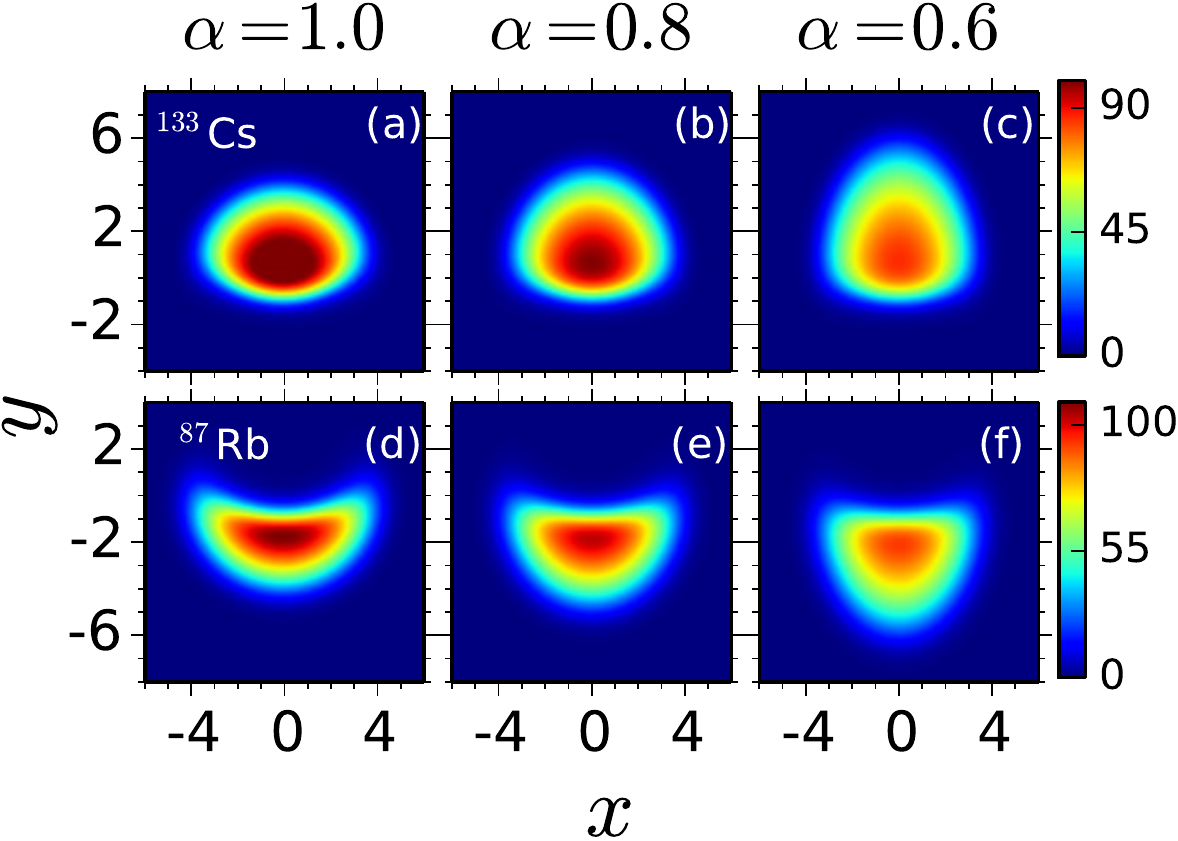}
    \caption{(Color online) Equilibrium condensate density profiles of 
$^{133}$Cs~-$^{87}$Rb TBEC for $a_{12} = 220a_0$ at zero temperature. (a)-(c)
Equilibrium density profile corresponding to $^{85}$Rb for three different
values of $\alpha$ ($\alpha$ = 1.0, 0.8, 0.5 (d)-(f)Equilibrium density
profile corresponding to $^{87}$Rb for the same three values of $\alpha$. The
color bar shows the value of $n_c$ measured in units of $a_{\rm osc}^{-2}$.
Spatial coordinates $x$ and $y$ are measured in units of $a_{\rm osc}$.}
    \label{denCsRb}
\end{figure}

The equilibrium density profiles for side by side density configuration are
shown in Fig.~\ref{denCsRb} for different values of $\alpha$. Unlike the 
general trend of no preferred orientation in the side-by-side phase segregation
with radially symmetric confining potential, in this case the phase 
segregation occurs along $y$ direction. This is appropriate since 
$\alpha < 1.0$ indicates $\omega_y < \omega_x$ which in turn sets $y$-axis 
as the preferred direction for the condensate to expand more freely and at 
appropriate parameter regime, the phase segregation occurs along $y$. The
orientation of the interface along $y$ direction occurs with the 
introduction of pinstripe, however small it may be.

In terms of the mode evolution, the immediate consequences of phase 
segregation along $y$ direction is the hardening of the zero energy mode
which emerges at phase separation. We observe that with the decrease in 
$\alpha$, the new zero energy mode regains energy and this can be attributed to 
the anisotropy induced stronger segregation along $y$ direction. In our earlier
work, in a different system we had reported the observation of the hardening 
of third Goldstone mode in TBEC that emerges at phase-separation, when the 
confining potentials have separated trap centers \cite{roy_15}.

\subsection{Excitations and dispersion relations for $^{133}$Cs~-$^{87}$Rb TBEC}
\label{cs-disp}

\begin{figure}[H]
 \includegraphics[width=8.5cm]{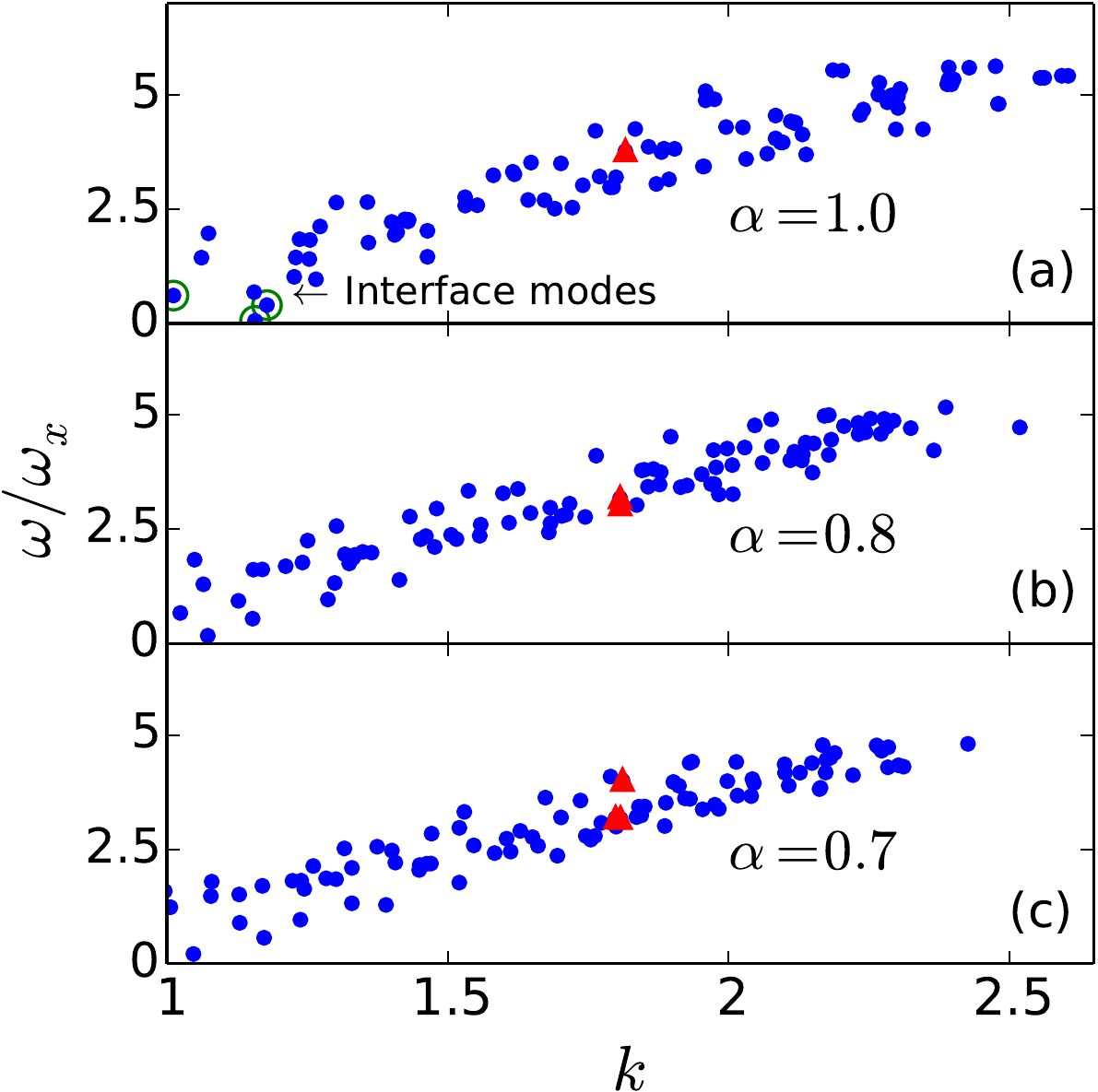}
    \caption{(Color online) (a) The dispersion relation for 
$^{133}$Cs~-$^{87}$Rb TBEC in immiscible (side by side) domain at 
$a_{12}$=220$a_0$ when $\alpha = 1.0$. In this plot the dispersion curves can
be broadly classified into two largely separated branches. The quasi particle
amplitudes corresponding to the upper branch (near energy 6.5$\hbar\omega_x$)
is shown in Fig.~\ref{qamp1-disprb}. For the lower branch (around energy range
1.7-2.7$\hbar\omega_x$), the quasi particle amplitudes is shown in 
Fig.~\ref{qamp2-disprb}. (b) The dispersion relation for immiscible 
(shell structured) $^{133}$Cs~-$^{87}$Rb TBEC at $a_{12}$=220$a_0$ when 
$\alpha = 0.8$. (c) The dispersion relation for immiscible (shell structured)
$^{133}$Cs~-$^{87}$Rb TBEC at $a_{12}$=220$a_0$ when $\alpha = 0.7$. As value
of $\alpha$ is reduced from 1.0 i.e., radial anisotropy makes this two
branches comparatively closer to each other.}
 \label{disp-csrb}
\end{figure}

\begin{figure}[H]
\centering
 \includegraphics[width=9.0cm]{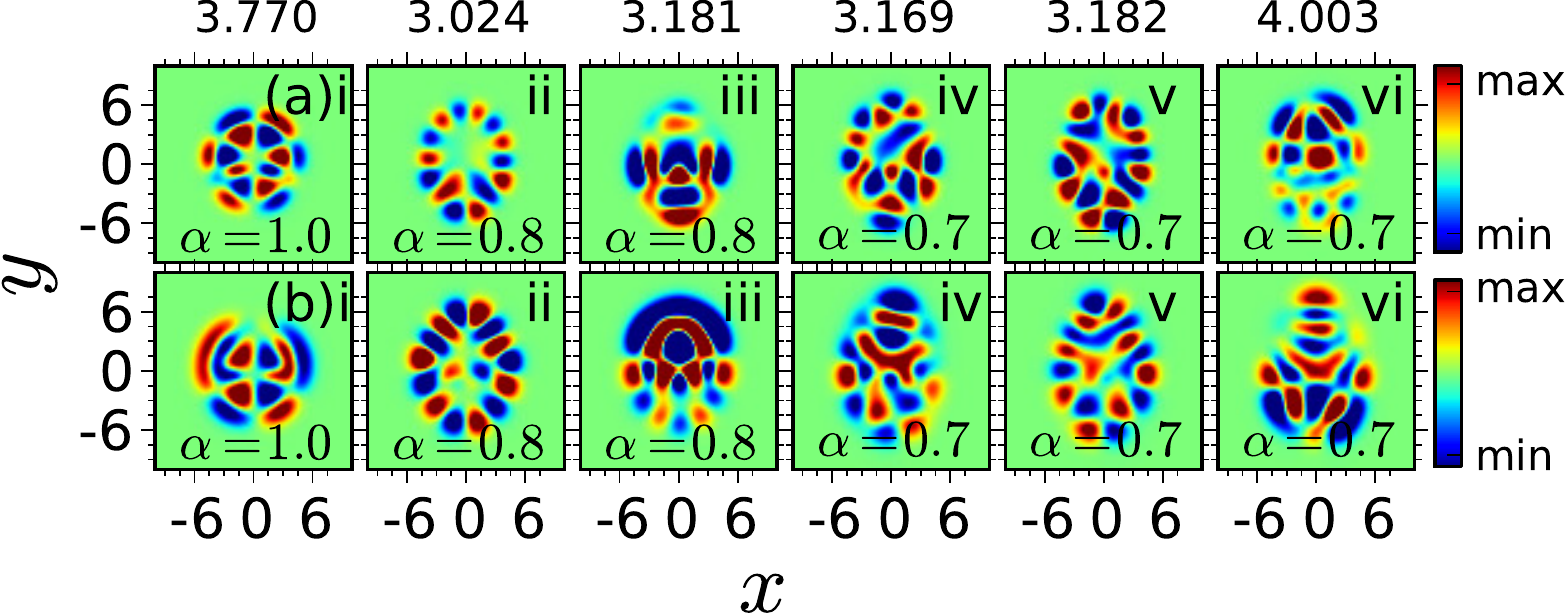}
    \caption{(Color online) Quasiparticle amplitudes for the immiscible
             (side by side) $^{133}$Cs~-$^{87}$Rb TBEC in space. 
             (a)i - vi: Quasi particle amplitudes for species $^{133}$Cs for a
             fixed value of $k \simeq 1.8$. (b)i - vi: Quasiparticle amplitudes
             for
             species $^{87}$Rb for the same value of $k$. In
             Fig.~\ref{disp-rbrb} this points are marked
             with red solid triangles for $k = 1.8$. Excitation energies
             corresponding to each set of $u_1(x,y)$ and $u_2(x,y)$ are
             mentioned at the top of each column. Here $u$s are in units of 
             $a_{\rm osc}^{-1}$. spatial coordinates $x$ and $y$ are measured in
             units of $a_{\rm osc}$.}
 \label{qamp-csrb}
\end{figure}

The dispersion curves for the $^{133}$Cs~-$^{87}$Rb TBEC in the immiscible
domain with side-by-side configuration are shown in Fig.~\ref{disp-csrb}. 
A prominent feature of the curves is that these are devoid of any discernible
trends, and this is due to the lack of rotational, reflection or scaling
symmetries. Some of the low-energy interface modes are identified by green 
circles and these are in agreement with the interface modes reported in 
Ref. \cite{ticknor_14} for TBEC with side by side configuration. To study the 
modes for specific $k$, we consider the modes at $k \simeq 1.8$ denoted by red 
solid triangles in Fig.~\ref{disp-csrb} and the changes in mode structures are
shown as function of $\alpha$ in Fig.~\ref{qamp-csrb}.

\begin{figure}[H]
\centering
 \includegraphics[width=9.0cm]{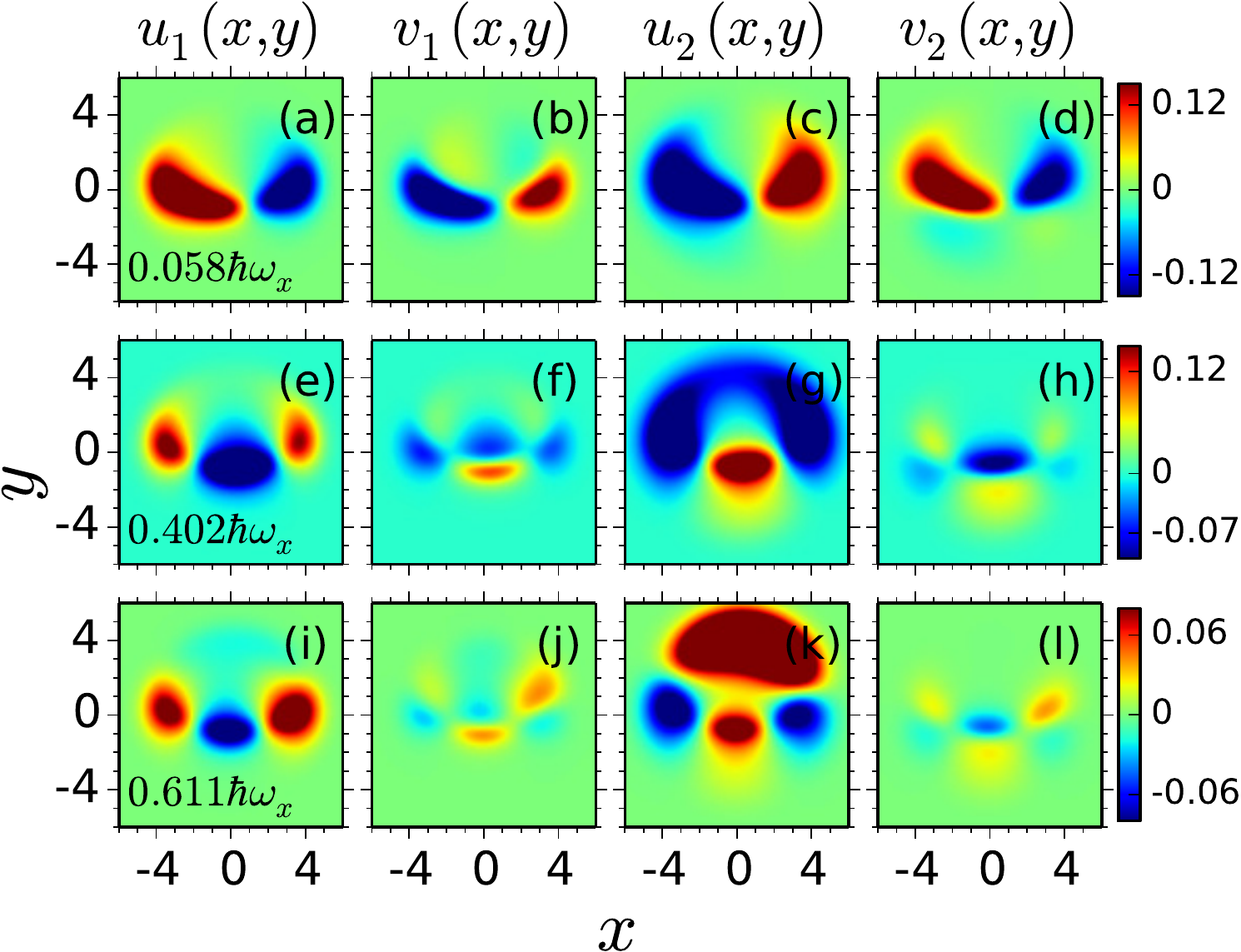}
    \caption{(Color online) Quasiparticle amplitudes corresponding to the
             interface modes marked as the green empty circle in 
             Fig.~\ref{disp-csrb} of the immiscible
             (side by side) $^{133}$Cs~-$^{87}$Rb TBEC for $\alpha = 1.0$ . 
             (a) - (d): Quasiparticles for the interface modes with energy
              0.058$\hbar\omega_x$.(e)-(h): Quasiparticles for the interface
              modes with energy 0.402$\hbar\omega_x$ and (i)-(l) is those with
              energy 0.611$\hbar\omega_x$. Here $u$s are in units of 
             $a_{\rm osc}^{-1}$. spatial coordinates $x$ and $y$ are measured in
             units of $a_{\rm osc}$.}
 \label{int-csrb}
\end{figure}

Fig.~\ref{int-csrb} corresponds to the localized excitation modes of the
segregated TBEC. These out-of-phase quasiparticle amplitudes describe the
interface excitations and localized only at the interface separating the two
condensates. The strength of quasiparticles $u_1(x,y)$ and $u_2(x,y)$ are large
compared to the quasiholes denoted by $v_1(x,y)$ and $v_2(x,y)$.


\section{Conclusions}
\label{conc}
In conclusion, we have characterized the low energy excitations of phase
segregated TBEC in presence of radial anisotropy with the Bogoliubov–de 
Gennes approach. For immiscible TBEC having shell
structured density profile, we have observed that the introduction of radial
anisotropy modifies the structure of the interface from circular to planar.
Our studies on $^{85}$Rb~-$^{87}$Rb TBEC, as an example of shell
structured density, shows that the interface and bulk modes have different 
dispersion relations in the rotationally symmetric geometry. However, 
anisotropy tends to merge the the dispersion relations. For the side by side 
geometry the effect of the radial anisotropy manifest through the breaking of 
rotational symmetry and interface orients along minor axis. This follows from
energy minimization through shorter interface geometry. For this case we
have chosen $^{133}$Cs~-$^{87}$Rb TBEC as an example and demonstrate that
the lack of symmetry lead to a dispersion curve which is devoid of any 
discernible trends. Furthermore, the effect of the anisotropy on the structure 
of the quasiparticles for these two systems are examined. One important
difference between the two in the immiscible domain is, the TBEC with 
side by side geometry has interface modes  as the lowest-energy modes. Whereas 
for the shell structured density profile the interface modes relatively 
higher energies.


\begin{acknowledgments}

The authors thank K. Suthar, S. Bandyopadhyay and R. Bai for useful
discussions. The results presented in the paper are based on the computations
using Vikram-100, the 100TFLOP HPC Cluster at Physical Research Laboratory, 
Ahmedabad, India.

\end{acknowledgments}

\appendix

\section{}
\begin{figure}[H]
 \includegraphics[width=9.0cm]{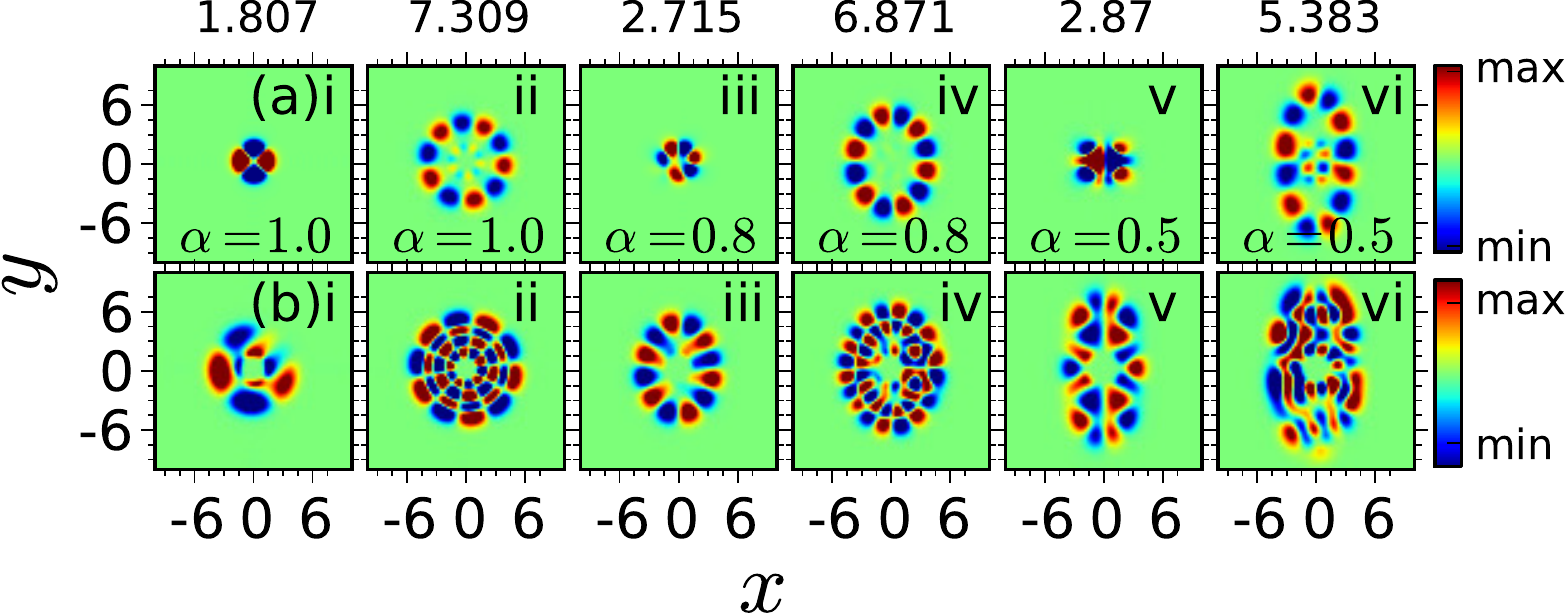}
    \caption{(Color online) Quasiparticle amplitudes for the immiscible
             (shell structured) $^{85}$Rb~-$^{87}$Rb TBEC in space for a
             particular $k \sim 1.85$. The corresponding mode energies are
             marked with maroon triangle 
             $({\color{maroon(html/css)}\blacktriangle})$ in
             Fig.~\ref{disp-rbrb}. (a)i- vi: Quasi particle amplitudes for
             species $^{85}$Rb. 
             (b)i - vi: Quasiparticle amplitudes for
             species $^{87}$Rb. (ii),(iv),(vi) in both the row (a) and (b)
             represent the interface modes. Whereas (i),(iii),(v) in row (a)
             denote the bulk excitations for $^{85}$ Rb and those in row (b)
             are bulk excitations of $^{87}$Rb. Each set of
             quasiparticles start from $\alpha = 1.0$ and move towards
             $\alpha = 0.5$. Excitation energy
             corresponding to each set of $u_1(x,y)$ and $u_2(x,y)$ are
             mentioned at the top of each column. Here $u$s are in units of 
             $a_{\rm osc}^{-1}$. spatial coordinates $x$ and $y$ are measured
             in units of $a_{\rm osc}$.}
 \label{qamp-k1p8}
\end{figure}

  To compare the structure of the interface and bulk modes at higher momenta,
we consider modes with $k \sim 1.85$ and plot of the quasiparticle amplitudes
are shown in Fig.~\ref{qamp-k1p8}. The plots correspond to the modes marked with
maroon triangle $({\color{maroon(html/css)}\blacktriangle})$ in the dispersion
curve in Fig.~\ref{disp-rbrb}. In Fig.~\ref{qamp-k1p8}, the plots in the 
top (bottom) row are the quasiparticles for $^{85}$Rb ($^{87}$Rb). Among these
the plots labeled with even number (ii, iv, and vi) are the interface modes and
the other are the bulk modes. Comparing the energies, top label in the plots,
the interface modes have energies which are more than double of the bulk modes 
with similar $k$, and this is due to the localized nature of the interface
modes. The important trend discernible is the decrease in mode energies,
for the same $k$, with the decrease in $\alpha$. This is on account of the 
reduced trapping frequency along the $y$-axis. The other effect of the 
anisotropy is the deformation of the interface quasiparticle particle when
the interface changes from circular to planer geometry. The appearance of the 
nonzero values in the central region in Fig.~\ref{qamp-k1p8}(a)(vi) is a 
The other noticeable trend is the drastic transformation in the quasiparticle 
amplitudes of $^{87}$Rb, which lies at the periphery, with decrease of 
$\alpha$. Compared to which the quasiparticle amplitudes of $^{85}$Rb,
which occupies the core region, the changes in their structure is not large.
\begin{figure}[t]
 \includegraphics[width=8.0cm]{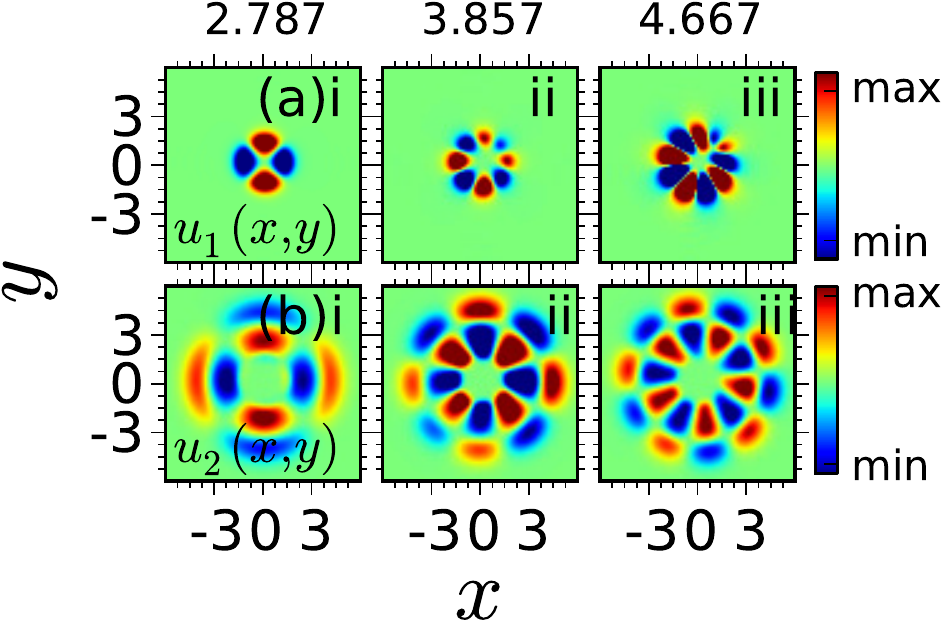}
    \caption{(Color online) Quasiparticle amplitudes corresponding to bulk
             excitations with relatively large-$k$ for the immiscible 
             (shell structured) $^{85}$Rb~-$^{87}$Rb TBEC in space for 
             $\alpha = 1.0$. (a)i - iii: Quasi particle amplitudes for species
             $^{85}$Rb which are composed of one radial nodes and $m$ varying
             with mode energies. (b)i - iii: Quasiparticle
             amplitudes for species $^{87}$Rb with two radial nodes and $m$
             varying with mode energies like other species. 
             In dispersion curves shown in Fig.~\ref{disp-rbrb}(a), these
             modes are marked with orange square 
             $({\color{orange(colorwheel)}\blacksquare})$. 
             Excitation energy for each set of $u_1(x,y)$ and $v_1(x,y)$ are
             highlighted at the top of the $u_1(x,y)$
             Here $u_1$s and $v_1$s are in units of $a_{\rm osc}^{-1}$. Spatial
             coordinates $x$ and $y$ are measured in units of $a_{\rm osc}$.}
 \label{qamp3-disprb}
\end{figure}

 A selected set of higher energy bulk modes, lower branch in the dispersion
curve, with larger number of radial nodes are shown in Fig.~\ref{qamp3-disprb}. 
The modes chosen are marked with orange square 
$({\color{orange(colorwheel)}\blacksquare})$ in the dispersion curve
which is shown in Fig.~\ref{disp-rbrb}. Like in the previous figures, 
$u_1(x,y)$ denotes the quasiparticles for $^{85}$Rb atoms and these are
given in Fig.~\ref{disp-rbrb}(a)(i-iii). Similarly, $u_2(x,y)$ correspond to 
those for $^{87}$Rb and are shown in \ref{disp-rbrb}(b)(i-iii). The notable 
feature is that the quasiparticles of $^{85}$Rb have only one radial node 
whereas those of $^{87}$Rb have two radial nodes.

\bibliography{mode}{}
\bibliographystyle{apsrev4-1}

\end{document}